# From planar junction to local junction: A new structure design of amorphous/crystalline silicon heterojunction solar cells for high efficiency and low cost

Haibin Huang[1], Lang Zhou[1,*], Jiren Yuan[1,2,*], and Zhijue Quan[3]

[1] *Institute of Photovoltaics, Nanchang University, Nanchang 330031, China*

[2] *Department of Physics, Nanchang University, Nanchang 330031, China*

[3] *National Institute of LED on Si Substrate, Nanchang University, Nanchang 330096, China*

*Corresponding authors. E-mails: lzhou@ncu.edu.cn; yuanjiren@ncu.edu.cn

**Abstracts:**

In order to obtain higher conversion efficiency and to reduce production cost for amorphous silicon/crystalline silicon (a-Si/c-Si) based heterojunction solar cells, a Heterojunction of Amorphous silicon and Crystalline silicon with Localized pn structure (HACL) has been designed. The potential performance of the HACL solar cell has been assessed by ATLAS simulation program. Its potential performance is compared with that of the Heterojunction with Intrinsic Thin film (HIT) and Heterojunction of Amorphous silicon and Crystalline silicon with Diffused junction (HACD) solar cells. The simulated results indicated that the conversion efficiency and the short-circuit current density of the HACL cell can reach to 28.18% and 43.06 mA/cm$^2$, respectively, and are higher than that of the HIT and HACD cells. The main reasons for the great improvement are (1) to increase the light utilization rate on both sides of the cell and (2) to enhance the collection efficiency of the photocarriers. Moreover, the HACL structure can decrease the consumption of rare materials, such as indium, since the transparent conductive oxide (TCO) can be free in this structure. It is concluded that the HACL solar cell is a promising structure for high efficiency and low cost.

## 1. Introduction

In the past few decades, amorphous silicon/crystalline silicon (a-Si:H/c-Si) based heterojunction solar cells have been attractive since they possess high open-circuit voltage ($V_{oc}$), low-cost fabrication process and low temperature coefficient [1-5]. In 1992, Sanyo (now Panasonic) company developed a heterojunction structure compose of a-Si:H and c-Si and called the HIT (Heterojunction with Intrinsic Thin-layer) solar cell [6]. The biggest highlight for the structure is that an intrinsic amorphous silicon (i-a-Si:H) thin-layer is introduced between the doped a-Si:H film and the c-Si base so that the carriers recombination near the interface can markedly reduced. The backward current density was only ~$10^{-8}$ A/cm$^2$ for this structure and a conversion efficiency of 18.1% for 1 cm$^2$ area was attained. In 2003, a symmetrical structure for HIT solar cells were designed by Sanyo [7]. The optical absorber is a textured n-type c-Si wafer, and the illuminated side and the back side are p/i a-Si:H film and i/n a-Si:H film, respectively. The main advantage of the symmetrical structure is the suppression of thermal and mechanical stress. The maximum conversion efficiency reached to 21.3% for 100-cm$^2$-area cell. Then, a conversion efficiency 22.3% was achieved in 2009 by improving the a-Si:H/c-Si heterojunction, improving the grid electrode and reducing the absorption in the a-Si:H and the TCO [8]. Particularly in 2013, a high conversion efficiency of 24.7% for 101.8 cm$^2$ area was achieved due to the improvement of TCO materials [9], the optimization of the Ag grid electrode for suppressing the shadow and resistance loss and the optimization of the resistance loss for i-a-Si:H layers.

In 2017, in order to further increase cell efficiency and reduce fabrication cost for a-Si:H/c-Si based heterojunction solar cells, we designed a novel bifacial silicon heterojunction structure of Ag grid/SiN$_x$/n$^+$-c-Si/n-c-Si/i-a-Si:H/p$^+$-a-Si:H/TCO/Ag grid (HACD cell) [10,11]. This structure and its fabrication process recipe proposed by our team have the advantages as follows: (1) it can remarkably reduce optical absorption loss at the illumined front of the cell since the doped and intrinsic a-Si:H layers, which possess very high absorption coefficient, are not employed at the front side of the cell; (2) it can reduce series resistance because high temperature fired Ag grids replaced low temperature solidified Ag grids and an intrinsic a-Si:H layer, which has high resistivity, is removed in the HACD structure; (3) a half of Indium Tin Oxides (ITO) materials can be saved. It is worth noting that the ITO material is very expensive, accounting for about 10% of the cost of the raw materials for a piece of HIT cell; (4) the fabrication processes are compatible with normal homojunction c-Si production lines, and the investment cost of new production line of the HACD solar cells could be remarkably cheap due to the special fabrication recipe design based on industrial processes.

Even so, we think that the above HACD cell structure is imperfect. First, it is still a planar junction, i.e., the emitter and back-field layer cover the entire surface. In this case, the surface area of the silicon wafer was covered by heavily doped layer, but the surface area of the silicon wafer is also the region where the photocarriers are most generated. This should bring about serious absorption loss. Second, those photocarriers in TCO layer and the heavily doped c-Si layer still need to be long-distance laterally transported to the grid line. It will result in a more recombination for the photocarriers. Third, there are large optical absorption loss due to the a-Si:H and TCO. In order to overcome these issues, we propose a bifacial heterojunction of amorphous/crystalline silicon with local pn structure (HACL cell) [12,13]. The primary difference between the HACL cell and the HACD/HIT cell is that all pn junctions are local for the HACL cell. The local pn junction can make the cell keep high open-circuit voltage, and also take full advantage of sunlight and thus enhance the photocurrent. Besides, the TCO layer, in which it contains rare and expensive indium, can be free for the HACL cell so that the cost will be reduced.

In this work, the potential conversion efficiency of the HIT, HACD and HACL cells is evaluated by the ATLAS program [14]. The device physics of the HACL cell is explored.

## 2. Device structure and simulation parameters

Figure 1 illustrates the schematic structure of the HIT, HACD and HACL solar cells used in this study. Fig. 1(a) shows the typical HIT solar cell structure, and Fig. 1(b) shows HACD solar cell structure which can be found detailed information in our previous work [10]. The HACL solar cell structure is shown in Fig. 1(c), in which the optical absorber is n-type c-Si. Each side of the optical absorber is constituted by emitter region and passivating region. Each emitter region comprises intrinsic and doped a-Si:H layers, whose width is about 50 μm, and the electrodes cover on the doped a-Si:H layers. Each passivating region includes heavy doped c-Si (very thin) and SiN$_x$ layers, whose width is about 1500 μm. Sunlight is incident from the passivating region into cell. The area of the emitter region covers about 3% of the area of the wafer, and the rest of the area of the wafer is covered by passivating region. The material parameters [10,15,16] used in this simulation are shown in Table 1. The simulated condition is under AM 1.5G, 100 mW/cm$^2$ and at 300 K.

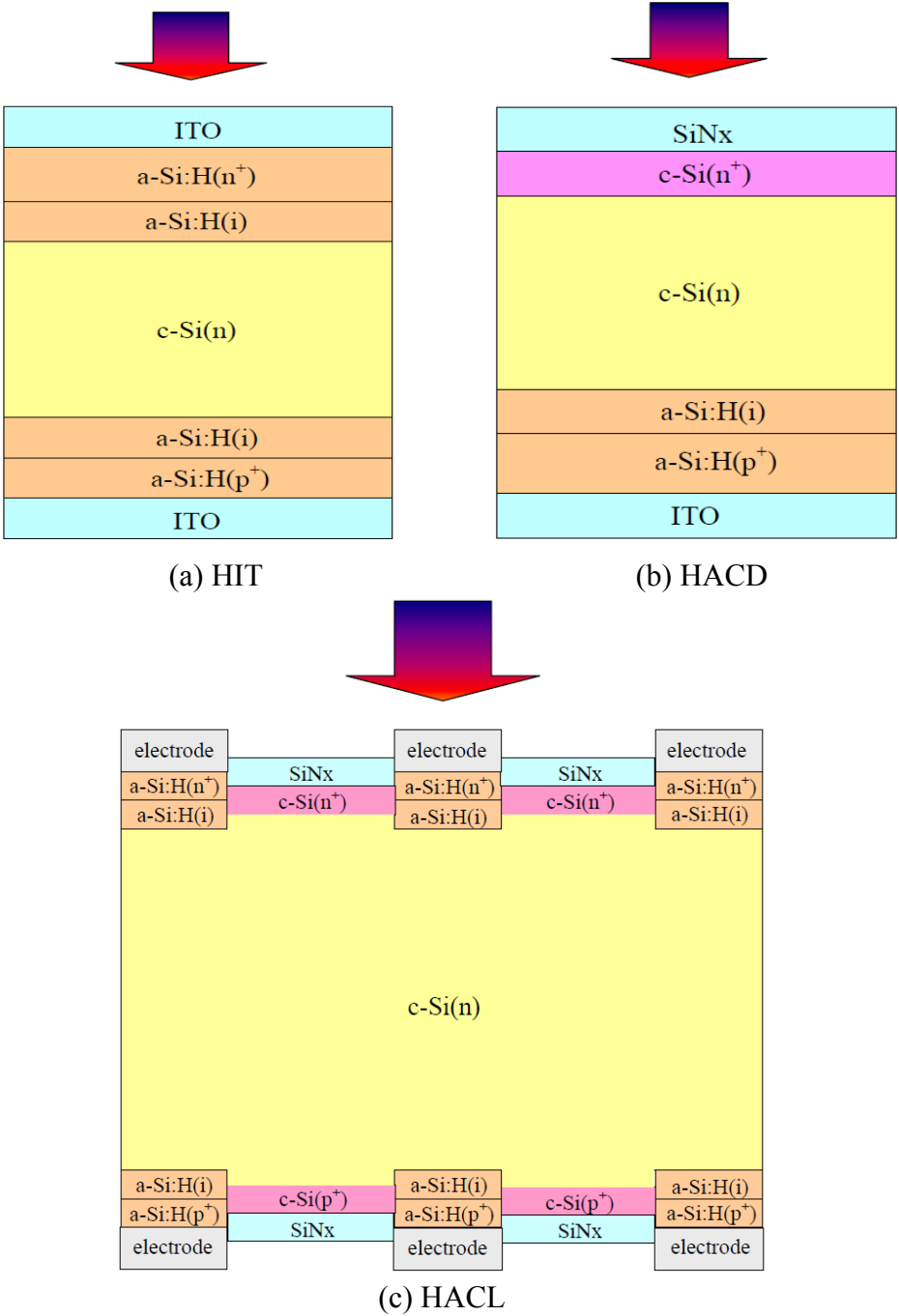

Fig. 1 Schematic of silicon heterojunction solar cells with different structures: (a) HIT, (b) HACD, and (c) HACL.

Table 1 Main parameters used for the simulation of HIT, HACD and HACL solar cells.

| Parameters and units | a-Si:H($p^+$) | a-Si:H(i) | a-Si:H($n^+$) | c-Si(n) | c-Si($n^+$/$p^+$) |
|---|---|---|---|---|---|
| Thickness (nm) | 10 | 10 | 10 | $1.5\times10^5$ | 20 |
| Electron affinity (eV) | 3.80 | 3.80 | 3.80 | 4.05 | 4.05 |
| Band gap (eV) | 1.72 | 1.72 | 1.72 | 1.12 | 1.12 |
| Relative dielectric constant | 11.90 | 11.90 | 11.90 | 11.90 | 11.90 |
| Effective conduction band density ($cm^{-3}$) | $2.50\times10^{20}$ | $2.50\times10^{20}$ | $2.50\times10^{20}$ | $2.80\times10^{19}$ | $2.80\times10^{19}$ |
| Effective valence band density ($cm^{-3}$) | $2.50\times10^{20}$ | $2.50\times10^{20}$ | $2.50\times10^{20}$ | $1.04\times10^{19}$ | $1.04\times10^{19}$ |
| Electron mobility ($cm^2V^{-1}s^{-1}$) | 10 | 20 | 10 | 1350 | 1350 |
| Hole mobility ($cm^2V^{-1}s^{-1}$) | 2 | 5 | 2 | 450 | 450 |
| Donor concentration ($cm^{-3}$) | 0 | 0 | $1\times10^{19}$ | $3\times10^{15}$ | $1\times10^{18}$ |
| Acceptor concentration ($cm^{-3}$) | $6\times10^{18}$ | 0 | 0 | 0 | 0 |
| Band tail density of states ($cm^{-3}eV^{-1}$) | $1\times10^{21}$ | $1\times10^{21}$ | $1\times10^{21}$ | $1\times10^{14}$ | $1\times10^{14}$ |
| Characteristic energy for donors, acceptors (eV) | 0.04, 0.06 | 0.04, 0.06 | 0.04, 0.06 | 0.01, 0.01 | 0.01, 0.01 |
| Capture cross-section for donor states, $e$, $h$[a] ($cm^2$)-Band tails | $1\times10^{-15}$, $1\times10^{-17}$ | $1\times10^{-15}$, $1\times10^{-17}$ | $1\times10^{-15}$, $1\times10^{-17}$ | $1\times10^{-15}$, $1\times10^{-17}$ | $1\times10^{-15}$, $1\times10^{-17}$ |
| Capture cross-section for acceptor states, $e$, $h$ ($cm^2$)-Band tails | $1\times10^{-17}$, $1\times10^{-15}$ | $1\times10^{-17}$, $1\times10^{-15}$ | $1\times10^{-17}$, $1\times10^{-15}$ | $1\times10^{-17}$, $1\times10^{-15}$ | $1\times10^{-17}$, $1\times10^{-15}$ |
| Gaussian density of states ($cm^{-3}$) | $7\times10^{18}$, $7\times10^{18}$ | $5\times10^{16}$, $5\times10^{16}$ | $1\times10^{18}$, $1\times10^{18}$ | – | – |
| Gaussian peak energy for donors, acceptor (eV) | 1.22, 1.22 | 1.22, 1.22 | 1.22, 1.22 | – | – |
| Standard deviation (eV) | 0.08 | 0.08 | 0.08 | – | – |
| Capture cross-section for donor states, $e$, $h$ ($cm^2$)-Midgap | $1\times10^{-14}$, $1\times10^{-15}$ | $1\times10^{-14}$, $1\times10^{-15}$ | $1\times10^{-14}$, $1\times10^{-15}$ | – | – |
| Capture cross-section for acceptor states, $e$, $h$ ($cm^2$)-Midgap | $1\times10^{-15}$, $1\times10^{-14}$ | $1\times10^{-15}$, $1\times10^{-14}$ | $1\times10^{-15}$, $1\times10^{-14}$ | – | – |
| Midgap density of states in c-Si ($cm^{-3}eV^{-1}$) | – | – | – | $1\times10^{12}$ | $1\times10^{12}$ |
| Switch-over energy (eV) | – | – | – | 0.56 | 0.56 |
| Capture cross-section for donor states, $e$, $h$ ($cm^2$) | – | – | – | $1\times10^{-14}$, $1\times10^{-15}$ | $1\times10^{-14}$, $1\times10^{-15}$ |
| Capture cross-section for acceptor states, $e$, $h$ ($cm^2$) | – | – | – | $1\times10^{-15}$, $1\times10^{-14}$ | $1\times10^{-15}$, $1\times10^{-14}$ |

[a] $e$, $h$ represents the electron and hole, respectively.

## 3. Results and Discussion

*3.1 Potential performance of HIT, HACD and HACL cells*

In order to verify the advantage of the HACL structure, potential performances of the HIT, HACD and HACL cells are calculated by ATLAS program. Fig. 2 shows the simulated results of the potential performances of the HIT, HACD and HACL cells. It is found that the HACL cell has the highest conversion efficiency. If sunlight is incident from the top side of the cell, a conversion efficiency of 27.84% for HACL cell can be achieved, while for the HACD and HIT cells, their conversion efficiencies are 27.35% and 26.58%, respectively. When the illumination enter from bottom side of the cell, the potential conversion efficiencies of the HIT, HACD and HACL cells are 26.53%, 27.03% and 28.18%, respectively. It is worth noting that the difference of conversion efficiency of three cells mainly depends on the short-circuit current density ($J_{sc}$). It can be seen from Fig. 2(a) that the $J_{sc}$ of the HACD cell is higher than that of the HIT cell. It is because the doped and intrinsic a-Si:H layers in the HIT cell bring about much optical absorption loss. For the HACL cell, its high $J_{sc}$ may be attributed to two aspects. One is that the area of the doped and intrinsic a-Si:H layers just account for about 3% of the surface of the c-Si base. So, more photons can be entered into the cell absorber to generate photocarriers. Another is that the unique HACL structure can help photocarriers to be better transported and collected. In order to illustrate the argument, the vector of electric field for the top part of the HACL cell is plotted as shown in Fig. 3a. It is observed that the electric field direction in emitter region is from $n^+$-a-Si:H to i-a-Si:H, then to n-c-Si absorber, and that in passivating region is from $n^+$-c-Si to n-c-Si. So, a large number of photocarriers generated at the c-Si wafer surface are driven into cell inside, and then they are transported to local pn junction and collected by electrodes (on the local pn junction). This process can avoid lateral transport for photocarriers so that it can improve collection efficiency of photocarriers. For traditional planar junction, lateral transport of photocarriers generated at the surface of the c-Si base can result in more recombination since there is heavy doping in emitter layer. Fig. 3b shows the electric field of the bottom part of the HACL cell. From the contours of electric field, we can obtain the same conclusion. As a result, one of the advantages for the local pn junction is that collection efficiency of photocarriers can be improved.

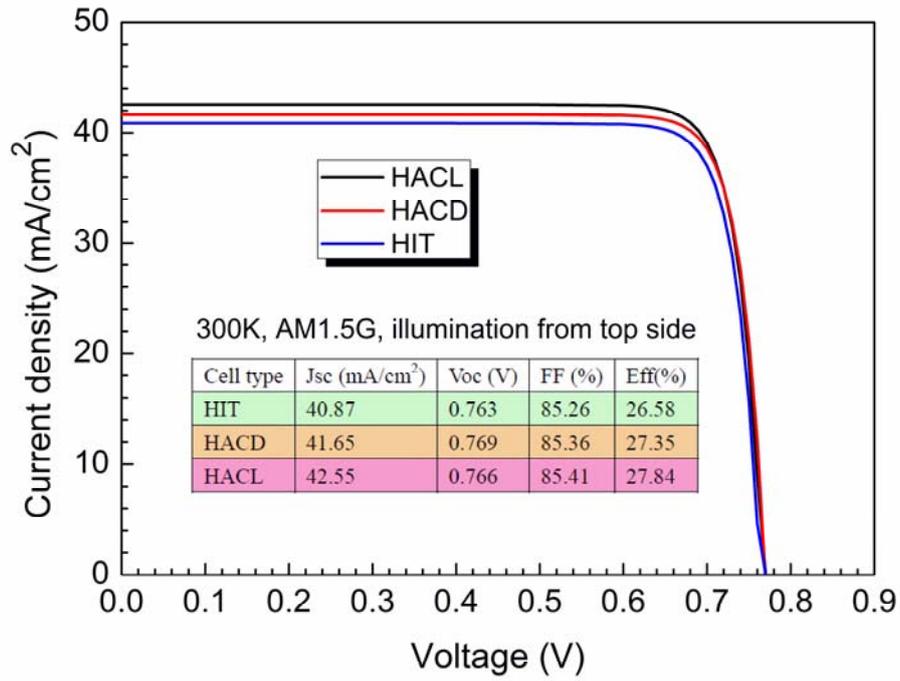

(a)

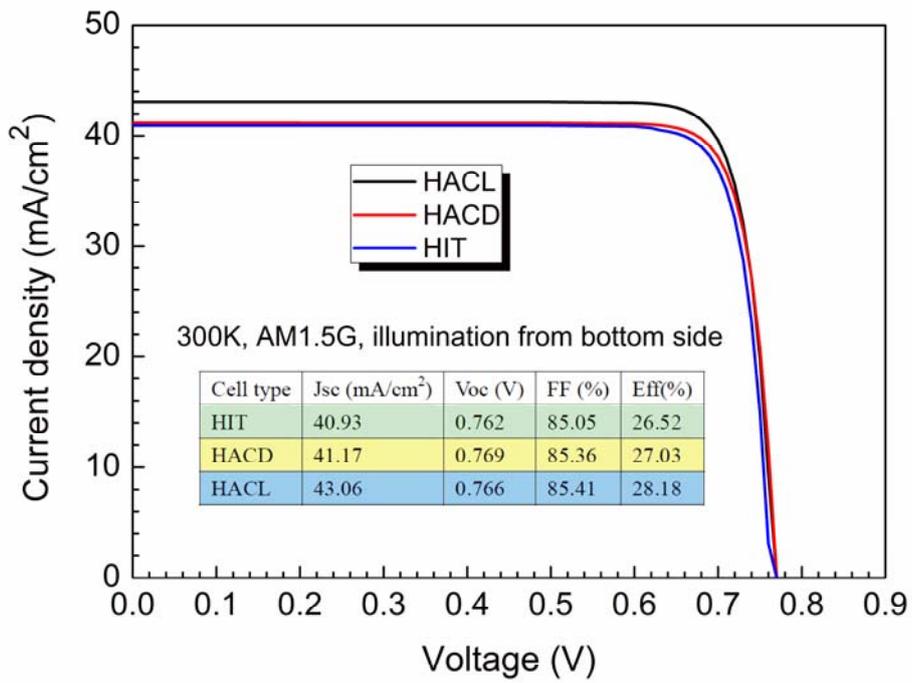

(b)

Fig. 2 Potential performances of the HIT, HACD and HACL cells, (a) illumination from top side; (b) illumination from bottom side.

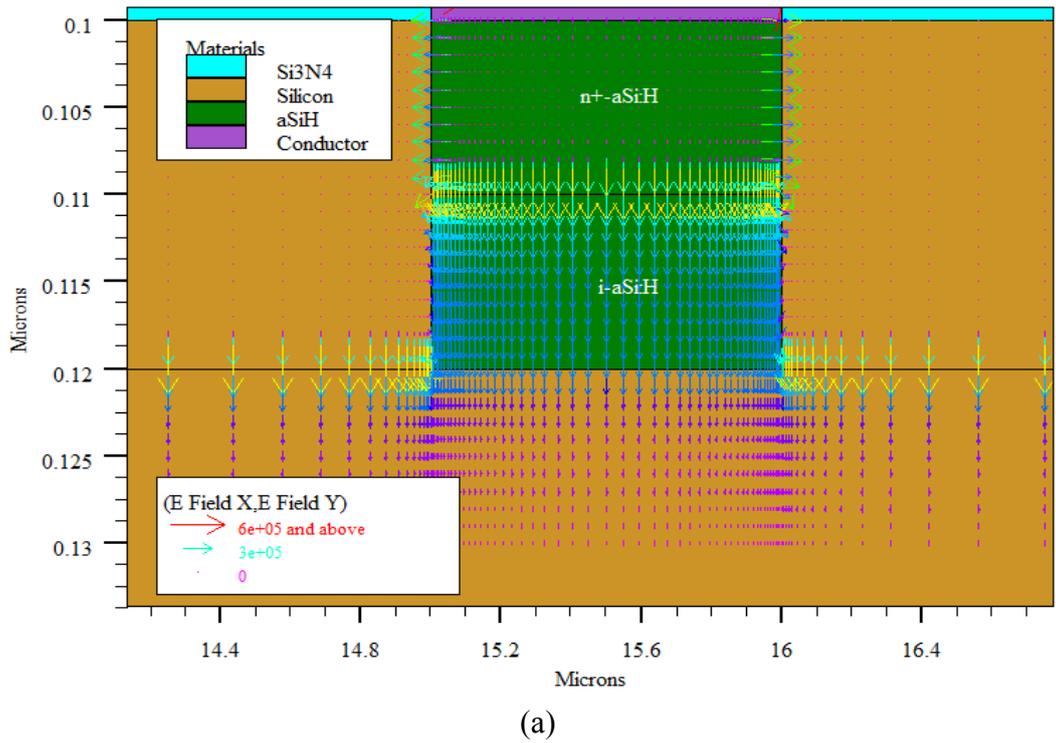

(a)

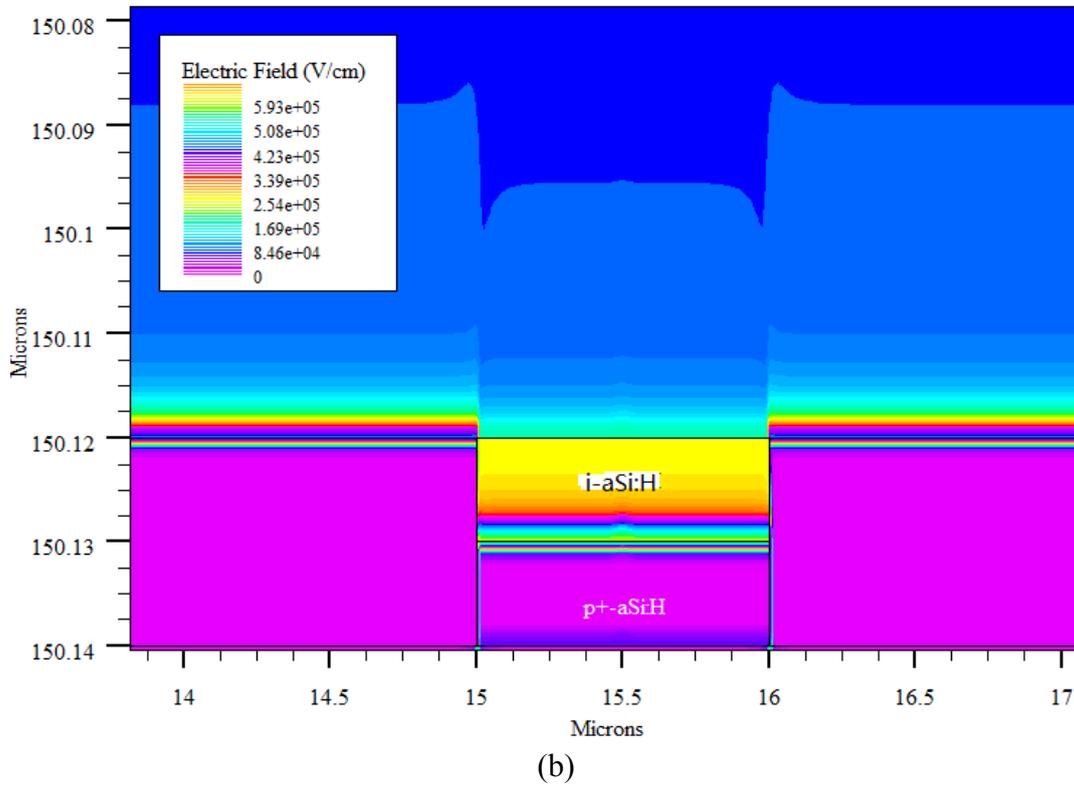

(b)

Fig. 3 Contours of electric field of the HACL cell, (a) from the top part (b) from the bottom part

*3.2 Effect of all n-type c-Si passivating region on HACL cell*

It can be seen from Fig. 1(c) that the passivating regions are covered by the n-type (top side) or p-type c-Si layers (bottom side) for the HACL cell. This design will bring about more steps for the production process. If the p-type c-Si passivating layer (bottom side) is replace by the n-type passivating one, i.e., the passivating layers are all n-type c-Si, it will be convenient to the double-sides diffusion process for forming double-sides n-type c-Si passivating layers. In order to prove its feasibility, the photovoltaic performances of two kinds of HACL cells ( cell-I and cell-II) were calculated. The cell with n-type and p-type c-Si passivating layers is cell-I, and that with all n-type c-Si passivating layers is cell-II. The calculated results are as listed in Table 2. It is found that the photovoltaic performances of cell-I and cell-II are very close. So, we can infer that the HACL cell with all n-type c-Si passivating layers can work well. But, it should be noticed that n-type c-Si passivating region at the bottom must be not connected with i-a-Si:H and $p^+$- a-Si:H layers in the emitter region. The gap between passivating region and emitter region can help to more effectively cut off the lateral transport of photocarriers. Moreover, the lateral pn junction at the bottom can be avoided to form due to the existance of the gap. In the fabrication process, the gap can be filled by insulator, such as $SiN_x$. Of course, it is good choice that if there is a gap between passivating region and emitter region at the top side of the HACL cell. The gap can also cut off the lateral transport of photocarriers. Our calculated results indicated that the effect of the gap at the top can slightly improve the cell performance (not shown here).

Table 2 Calculated results of photovoltaic performance of the cell-I and the cell-II.

| Cell types | Jsc (mA/cm$^2$) | Voc (V) | FF (%) | Eff(%) |
|---|---|---|---|---|
| I | 42.55 | 0.766 | 85.43 | 27.84 |
| II | 42.21 | 0.766 | 85.28 | 27.58 |

**4. Conclusions**

The potential performances of the HIT, HACD and HACL cells have been evaluated by using the ATLAS simulation program. The HACL structure possesses the highest conversion efficiency (28.18%) and short-circuit current density (43.06 mA/cm$^2$). The excellent performance for the HACL cell should be originated from the unique structure design based on local pn junction. The advantage of the new structure is that more photons can be entered into the cell absorber to generate photocarriers since a-Si:H materials are free in the non-junction regions with big area and the collection efficiency of photocarriers can be improved since the lateral transport of photocarriers is avoided. Moreover, the TCO materials, in which there is rare and valuable metal element of indium, are free for the HACL structure so that the production cost can be reduced. The results show that the HACL structure has great potential for high efficiency and low cost.

**Acknowledgements**

This work was supported by the National Natural Science Foundation of China (Grant Nos. 61464007 and 61741404), the Natural Science Foundation of Jiangxi Province of China (Grant No. 20181BAB202027).

**References**

[1] Y. Yao, X. Xu, X. Zhang, H. Zhou, X. Gu, S. Xiao, Enhanced efficiency in bifacial HIT solar cells by gradient doping with AFORS-HET simulation, Materials Science in Semiconductor Processing, 77, 16 (2018).

[2] Z. Wu, L. Zhang, R. Chen, W. Liu, Z. Li, F. Meng, Z. Liu, Improved amorphous/crystalline silicon interface passivation for silicon heterojunction solar cells by hot-wire atomic hydrogen during doped a-Si:H deposition, Applied Surface Science, 475, 504 (2019).

[3] L. Zhao, G. Wang, H. Diao, W. Wang, Physical criteria for the interface passivation layer in hydrogenated amorphous/crystalline silicon heterojunction solar cell, Journal of Physics D: Applied Physics, 51, 045501 (2018).

[4] K. Yoshikawa, H. Kawasaki, W. Yoshida, T. Irie, K. Konishi, K. Nakano, T. Uto, D. Adachi, M. Kanematsu, H. Uzu, K. Yamamoto, Silicon heterojunction solar cell with interdigitated back contacts for a photoconversion efficiency over 26%, Nature Energy 2, 17032 (2017).

[5] M.A. Green, Y. Hishikawa, E.D. Dunlop, D.H. Levi, J. Hohl-Ebinger, A.W.Y. Ho-Baillie1, Solar cell efficiency tables (version 51), Prog. Photovolt. Res. Appl. 26, 3 (2018).

[6] M. Tanaka, M. Taguchi, T. Matsuyama, T. Sawada, S. Tsuda, S. Nakano, H. Hanafusa, Y. Kuwano, Jpn. J. Appl. Phys. 31, 3518 (1992).

[7] M. Tanaka, S. Okamoto, S. Tsuge, S, Kiyama, Development of HIT solar cells with more than 21% conversion efficiency and commercialization of highest performance hit modules, 3rd World Conference on Photovoltaic Energy Conversion, May 11-18, 2003, Osaka, Japan.

[8] Y. Tsunomura, Y. Yoshimine, M. Taguchi, T. Baba, T. Kinoshita, H. Kanno, H. Sakata, E. Maruyama, M. Tanaka, Twenty-two percent efficiency HIT solar cell, Solar Energy Materials and Solar Cells, 93, 670 (2009).

[9] M. Taguchi, A. Yano, S. Tohoda, K. Matsuyama, Y. Nakamura, T. Nishiwaki, K.Fujita, E. Maruyama, 24.7% Record Efficiency HIT Solar Cell on Thin Silicon Wafer, IEEE Journal of Photovoltaics, 4, 96 (2014).

[10] H.B. Huang, G.Y. Tian, L. Zhou, J.R. Yuan, W.R. Fahrner, W.B. Zhang, X.B. Li, W.H. Chen, R.Z. Liu, Simulation and experimental study of a novel bifacial structure of silicon heterojunction solar cell for high efficiency and low cost, Chinese Physics B, 27, 038502 (2018).

[11] H.B. Huang, L. Zhou, J.R. Yuan, C. Gao, Z.H. Yue, Crystalline silicon bifacial solar cell structure with HAC-D characteristics, CN108336156A, July, 27, 2018.

[12] H.B. Huang, L. Zhou, J.R. Yuan, C. Gao, Z.H. Yue, Local amorphous silicon emitting electrode crystalline silicon back surface field bifacial solar cell structure, CN108336157A, July, 27, 2018.

[13] J.R. Yuan, L. Zhou, H.B. Huang, C. Gao, Z.H. Yue, Si-based dual-faced solar cell structure based on local emitting electrode features, CN108461569A, Aug. 28, 2018.

[14] ATLAS User's Manual—device Simulation Software, SILVACO, Santa Clara, CA, 2016.

[15] J.R. Yuan, H.L. Shen, L.F. Lu, T.R. Wu, X.C. He, Simulation of HIT solar cells with μc-3C-SiC:H emitter, Optoelectronics and Advanced Materials-Rapid Communications, 4, 1211 (2010).

[16] N. Hernandez-Como, A. Morales-Acevedo, Simulation of hetero-junction silicon solar cells with AMPS-1D, Solar Energy Materials and Solar Cells, 94, 62 (2010).